\documentclass[prd,showpacs,superscriptaddress]{revtex4-1}
\usepackage{amsmath}
\usepackage{graphicx}
\usepackage{array}

\begin{document}
\title{Decay $t\to c\gamma$ in models with $SU_L(3)\times U_X(1)$ gauge symmetry}

\author{I. Cort\'es-Maldonado}
\affiliation{Departamento de F\'{\i}sica, CINVESTAV IPN,
Apartado Postal 14-740, 07000, M\'exico D. F., M\'exico.}
\author{G. Hern\'andez-Tom\'e}
\affiliation{Facultad de
Ciencias F\'\i sico Matem\'aticas, Benem\'erita Universidad
Aut\'onoma de Puebla, Apartado Postal 1152, Puebla, Pue., M\'
exico}
\author{G. Tavares-Velasco}
\email[Corresponding author: ]{gtv@fcfm.buap.mx}
\affiliation{Facultad de
Ciencias F\'\i sico Matem\'aticas, Benem\'erita Universidad
Aut\'onoma de Puebla, Apartado Postal 1152, Puebla, Pue., M\'
exico}

\begin{abstract}

The one-loop level mediated $t\to c \gamma$ decay is analyzed in the framework of 331 models, which
are based on the $SU_L(3)\times U_X(1)$ gauge symmetry and require that the quark families transform
differently in order to cancel anomalies, thereby inducing three-level flavor-changing neutral
currents mediated by  an extra neutral gauge boson, $Z'$, and  a neutral scalar boson,                                                            
$\phi$. These models also predict new charged gauge  and scalar bosons, together with three new
quarks, which can be  exotic (with  electric charges of $-4/3e$ and $5/3e$) or standard model like. Apart
from the contribution of the $W$ boson, the $t\to c\gamma$ decay receives contributions induced by
the extra gauge boson and the neutral scalar boson, which are generic for 331 models. In the
so-called minimal 331 model, there are additional  contributions from the new charged gauge and scalar
bosons accompanied by the exotic quarks. We present analytical results for the most general $t\to
c\gamma$ amplitude in terms of transcendental functions. For the numerical analysis we focus on the
minimal 331 model:  the current bounds on the model parameters are examined and a particular
scenario is discussed in which the corresponding branching ratio could be of the order of $10^{-6}$,
with the dominant contributions arising from the charged gauge bosons and a relatively light neutral
scalar boson with flavor-changing couplings, whereas the $Z'$ contribution would be of the order of
$10^{-9}$ for $m_{Z'}>2$ TeV.  However, a further suppression could be expected due to a potential
suppression of the values of the flavor-changing coupling constants. Under the same assumptions, in
331 models without exotic quarks, the $t\to c\gamma$ branching ratio would receive the dominant
contribution from the neutral scalar boson, which could be of the order of $10^{-7}$ for a Higgs
mass of a few hundreds of GeVs.

\end{abstract}
\pacs{14.65.Ha,12.60.Cn,14.80.-j}
\maketitle

\section{Introduction}

Despite  recent evidences of a neutral Higgs-like particle at the CERN LHC
\cite{ATLAS:2012ae,*Chatrchyan:2012tx}, it is  necessary to search for effects beyond the
standard model (SM) as there are still some open questions. Along these lines, several SM extensions
have
been proposed,
such as  two-Higgs doublet models (THDM) \cite{Haber:1978jt}, left-right symmetric models \cite{Mohapatra:1974gc}, supersymmetric models \cite{Haber:1984rc}, left-right supersymmetric models \cite{Aulakh:1998nn},  331 models \cite{Pisano:1991ee,Frampton:1992wt}, little Higgs models \cite{ArkaniHamed:2001nc,ArkaniHamed:2002qy}, and
extra dimension models \cite{Randall:1999ee}, just to mention some of the most popular ones. Such models predict new physics effects in the form
of new particles, corrections to the SM couplings or non-SM couplings.
Among the new predicted
particles there are, for instance,
exotic quarks, CP-even and CP-odd neutral scalar bosons, singly and doubly charged scalar bosons,
extra neutral  gauge bosons, singly and doubly charged gauge bosons, etc. It may be that there was
not enough energy to directly produce any of these hypothetical states  at particle colliders,  and
so  their only observable sign would arise indirectly via their loop effects. In
particular, the new
particles may give rise to sizeable effects on one-loop induced processes, such as the flavor
changing neutral current (FCNC) decays of the top quark $t\to cV$ ($V=\gamma, Z$). Due to its heavy
mass, it has been long conjectured that top quark physics offers an opportunity to test the SM and
search for new physics effects \cite{Chakraborty:2003iw}. The rate for the  decay $t\to c\gamma$ is
negligibly small in the SM due to the GIM mechanism: the respective branching fraction is of the
order of $10^{-10}$ \cite{DiazCruz:1989ub}. Since the sensitivity of ATLAS to the  $t\to c\gamma$
branching
ratio  at the  LHC is expected to be of the order of $10^{-4}$, it is worth studying such a process
in SM extensions, where its branching ratio can be enhanced by several orders of magnitude
\cite{Larios:2006pb}: this decay  has been studied, for instance, in  the two-Higgs doublet model
[${\rm BR}(t\to c\gamma)\sim 10^{-7}$] \cite{Eilam:1990zc,*Diaz:2001vj}, technicolor
[${\rm BR}(t\to c\gamma)\sim 10^{-7}$] \cite{Lu:1996ji,*Wang:1994qd},  topcolor assisted technicolor
[${\rm BR}(t\to c\gamma)\sim 10^{-7}$] \cite{Lu:1998gm,*Lu:2003yr}, supersymmetric models [${\rm
BR}(t\to c\gamma)\sim 10^{-6}-10^{-5}$] \cite{Couture:1994rr,*Li:1993mg,*Lopez:1997xv,Yang:1997dk},
left-right supersymmetrical models [${\rm BR}(t\to c\gamma)\sim 10^{-6}$] \cite{Frank:2005vd},
extra dimensions [${\rm BR}(t\to c\gamma))\sim 10^{-10}$] \cite{GonzalezSprinberg:2007zz},
models with an extra neutral gauge boson [${\rm BR}(t\to c\gamma))\sim 10^{-8}$]
\cite{CorderoCid:2005kp}, etc.  On the other hand, a model-independent analysis via the effective
Lagrangian approach \cite{Han:1996ep} put the upper constraint  ${\rm BR}(t\to c\gamma)\lesssim
10^{-2}$ by using the experimental bounds on the  $b\to s\gamma$ decay. Also, by means of the
effective Lagrangian approach, the contribution of a neutral scalar to ${\rm BR}(t\to c\gamma)$ was
found to be the order of $10^{-8}$ \cite{CorderoCid:2004vi}.

We will calculate  the $t\to c\gamma$ decay in the framework of  models based on the
$SU_c(3)\times SU_L(3)\times U_X(1)$ gauge symmetry, which for short are called 331 models and have
been the source of considerable attention in the literature. The idea of embedding the
$SU_L(2)\times U_Y(1)$
gauge group into $SU_L(3)\times U_X(1)$  in order to explain the observation of
neutrino-induced trimuon events \cite{Barish:1977bg} was discussed in \cite{Lee:1977qs}, though
similar models had already been conjectured \cite{Segre:1976rc,*Yoshimura:1976ex,*Fritzsch:1976dq}.
Although these models were soon ruled out, another SM extension based on the $SU_L(3)\times U_X(1)$
gauge group was proposed  by the authors of  Ref. \cite{Pisano:1991ee},  motivated
by the need of a doubly charged gauge boson to restore the  unitarity of the cross section of the
process $e^-e^-\to W^-V^-$. An almost identical model was proposed independently in
\cite{Frampton:1992wt}, but with a different motivation: the need of a chiral theory for doubly
charged gauge bosons. Such exotic particles had first been predicted in an $SU(15)$ grand unified
theory, which ensured proton-stability but required mirror fermions to cancel anomalies
\cite{Frampton:1989fu}. In 331 models, one fermion family must transform under the $SU_L(3)$ group
differently from the other two families in order to cancel  anomalies, thereby allowing for a
solution to the flavor problem: it is necessary that the number of fermion families is
a multiple of the quark color number. Also, if the third fermion family is the chosen one to
transform differently, 331 models may provide a hint for an eventual understanding of the heaviness
of the top
quark. Another appealing aspect of these models is that they can accommodate naturally the
Peccei-Quinn symmetry \cite{Pal:1994ba}. In the
phenomenological side, since the $SU_L(2)$ fermion doublets are promoted to $SU_L(3)$ triplets, 331
models  require new fermion particles. The way in which the fermion triplets are completed and the  chosen
$SU_L(3)\times U_X(1)$ representation for these triplets in order to cancel anomalies
give rise to distinct 331 model versions. In particular, the most popular ones  are
the minimal 331 model \cite{Pisano:1991ee,Frampton:1992wt} and
the 331 model with right-handed neutrinos
\cite{Foot:1994ym,Hoang:1995vq}. Other proposed 331 models can be found in Refs.
\cite{Pleitez:1992xh,Ozer:1995xi,Duong:1993zn,Montero:2000ng,Ferreira:2011hm}, and a general
treatment of  331 models without exotic quarks can be found in Refs.
\cite{Ponce:2001jn,Diaz:2004fs}. In addition, although with different purpose and  structure, a
little Higgs model with  global symmetry under the group $[SU(3)\times U(1)]^2$ and local symmetry
under the subgroup
$SU_L(3)\times U_X(1)$  was proposed in Ref. \cite{Schmaltz:2004de}, while its ultraviolet
completion
was studied in Ref. \cite{Kaplan:2004cr}. This model is an effective theory valid up to the scale of the
TeVs, which is known as the simplest little Higgs model and shares the same mechanism of
anomaly cancellation as that of the 331 model with right-handed neutrinos.

Apart from reproducing the SM, 331 models predict several new particles. In the gauge sector, the
typical signatures are an extra neutral gauge boson and a new singly charged gauge boson. Depending
on the particular version of the model, there could
be either a new doubly charged gauge boson, as in the minimal 331 model, or a new neutral no
self-conjugate gauge boson, as in the 331 model with right-handed neutrinos. As far as the scalar
sector is concerned, although the minimal 331 model requires three scalar triplets to accomplish the
spontaneous symmetry breaking (SSB) and a sextet to endow the leptons
with realistic masses, other versions require a more economical set of scalar multiplets. In this
sector there could be new neutral, singly, and doubly charged physical scalar bosons. In the quark
sector, three
new quarks must be introduced to complete the $SU_L(3)$ triplets: in the minimal 331 model there are
three new
exotic quarks, two of them have
electric charge of $-4/3e$,  while  the remaining one has charge of $5/3e$; however,
there
are
331 models in which the new quarks do not have exotic charges
\cite{Pleitez:1992xh,Foot:1994ym,Ozer:1995xi,Ponce:2001jn,Diaz:2004fs}. The fact that the fermion
families
transform differently under the gauge group gives rise to FCNC at the three level mediated by the
extra neutral gauge boson and the new neutral scalar bosons, which in turn can induce at the
one-loop level the $t\to
c\gamma$ decay, which can also be induced by the charged  gauge  and
scalar bosons. Below
we will calculate such a decay in the framework of 331 models and analyze
the
magnitude of the corresponding branching ratio considering the current constraints on the model
parameters from experimental data.

The rest of the presentation is organized as follows. In Section \ref{model} we present an overview
of 331 models and their potential sources of flavor change. Section \ref{calculation} is devoted to
the calculation of the $t\to c\gamma$ decay amplitude, while the numerical analysis and discussion
are
presented in Sec. \ref{analysis}. The conclusions and outlook are presented in Sec.
\ref{conclusions}.

\section{The model}
\label{model}

The motivation and general description of 331 models have already been discussed. We
turn to discuss briefly those aspects relevant for our calculation. In  331 models, the
charge operator is defined  by $Q=T^3+\beta T^8+ X $, where
$T^i=\lambda^{i}/2$, with $\lambda^i$ the Gell-Mann matrices and $X$ the $U_X(1)$ quantum number.
Specific values of $\beta$ give rise to distinct models
with a peculiar particle content: the minimal 331 model arises when $\beta=\pm\sqrt{3}$, in
which case there are three new exotic quarks and a new doubly charged gauge boson; when  $\beta=\pm
1/\sqrt{3}$, there are no exotic quarks but new SM-like quarks and  a   no
self-conjugate neutral gauge boson.

The generic contributions to the $t\to c\gamma$ decay in 331 models arise from the extra neutral
gauge boson and the neutral Higgs bosons. In this work  we will focus mainly on the minimal 331
model, which is the most popular
version of these models and the one that predicts additional contributions to the $t\to
c\gamma$ decay: those
mediated by the exotic quarks along with the new charged gauge and scalar bosons.
Nevertheless, our results will be rather
general and useful to estimate the size of the $t\to c\gamma$ branching ratio in other 331 models.

In the following, we will not discuss about the lepton sector as it is not relevant for the present
work.
In the quark sector,  three new quarks are required to complete the $SU_L(3)$
triplets.  In order to cancel anomalies, the first two quark families transform
under $SU_L(3)\times U_X(1)$ as follows:

\begin{equation}
Q_{1,2}=\left(\begin{array}{c}
u_{1,2}\\
d_{1,2}\\
D_{1,2}
\end{array}\right)
: (3,-1/3),\,\,
\begin{array}{c}
u_{1,2}^c: (1,-2/3),\\
d_{1,2}^c: (1,+1/3),\\
D_{1,2}^c: (1,+4/3),
\end{array}
\end{equation}
with $D_1=D$ and $D_2=S$. The numbers inside the parenthesis are the
$SU_L(3)\times
U_X(1)$ quantum numbers. On the other hand, the third quark family transforms as a triplet:

\begin{equation}
Q_3=\left(\begin{array}{c}
b\\
-t\\
T
\end{array}\right)
: (3^*,2/3),\,\,
\begin{array}{c}
b^c: (1,+1/3),\\
t^c: (1,-2/3),\\
T^c: (1,-5/3).
\end{array}
\end{equation}
As already mentioned, as a consequence of this representation, the new exotic
quarks have electrical charge of $Q_{D,S}=-4/3e$ and $Q_T=5/3e$.

The scalar sector of 331 models has been studied extensively
\cite{Tonasse:1996cx,Hoang:1997su,Tully:1998wa,Ponce:2002sg,Diaz:2003dk}. In the minimal
model, one triplet, $\phi_Y$, is necessary to break $SU_L(3)\times U_X(1)$ into $SU_L(2)
\times U_Y(1)$, and two  triplets, $\phi_{1,2}$,  are required to break $SU_L(2)\times U_Y(1)$
into $U_{em}(1)$. In addition, one scalar sextet, $H$, is required to give realistic masses to the
leptons. More recently, it has been noted that SSB can  be achieved with
only two scalar triplets \cite{Ferreira:2011hm}, but  the masses of the charged leptons must be
generated via nonrenormalizable effective operators.  On the contrary, in 331 models without exotic
charge quarks, a scalar sector with two or three scalar triplets is enough to achieve SSB and endow
all the particles with masses
\cite{Hoang:1997su,Ponce:2002sg,Diaz:2003dk}.

In the minimal 331 model, the scalar triplets
have the following quantum numbers:

\begin{equation}
\phi_Y= \left( \begin{array}{c}
\Phi_Y \\
\phi^0
\end{array} \right): (3,1),\quad
\phi_1= \left( \begin{array}{c}
\Phi_1 \\
\Delta^-
\end{array} \right): (3,0),\quad
\phi_2= \left( \begin{array}{c}
\widetilde{\Phi}_2 \\
\rho^{--}
\end{array} \right):  (3,-1),
\end{equation}
where $\Phi_i=(\phi^{+}_i,\phi^0_i)^T$ and
$\Phi_Y=(G_Y^{++},G_Y^+)^T$ are $SU_L(2)$ doublets with hypercharge 1  and 3, respectively, and
$\widetilde{\Phi}_i=i\,\tau^2\,\Phi^*_i$. Here
$G_Y^{++}$ and $G_Y^{+}$ are the would-be Goldstone bosons
associated with  new doubly  and singly charged gauge bosons, whereas the real
and imaginary parts of $\phi^0$ correspond to one physical Higgs boson and the would-be Goldstone
boson associated with an extra neutral gauge boson, respectively. Also,  $\Delta^-$
and $\rho^{--}$ are singlets of $SU(2)_L$ with hypercharge $-2$ and $-4$, respectively. As for the
scalar sextet, it has no significance for this work as it is only necessary to give realistic masses
to the leptons and so it does not couple to the quarks.

The covariant derivative in the fundamental representation of
$SU_L(3)\times U_X(1)$ can be written as

\begin{equation}
{D}_\mu=\partial_\mu+i\,g\,\frac{\lambda^a}{2}
\,W^a_\mu+i\,g_XX\,\frac{\lambda^9}{2}\,V_\mu, \qquad 
\end{equation}
where $a$ runs from 1 to 8, $W_\mu$ and $V_\mu$ are the $SU_L(3)$ and $U_X(1)$ gauge fields, and
$\lambda^9=\sqrt{2/3}\,{\rm diag}(1,1,1)$. By matching the gauge
coupling constants, it is found that $g_X=\sqrt{6}s_W/\sqrt{1-4s_W^2}$, with the usual short-hand
notation $s_W=\sin\theta_W$.

In the first stage of SSB, the vacuum expectation value (VEV) of the $\phi_Y$ triplet,
$\phi^\dag_{Y0}=(0,0,u/\sqrt{2})$, triggers the breaking of the
$SU_L(3)\times U_X(1)$ gauge group into
$SU_L(2) \times U_Y(1)$, thereby giving rise to two mass-degenerate singly and doubly charged
bosons, which are called bileptons as they carry two units of lepton number. They are given as
follows:
\begin{eqnarray}
Y^{-}_\mu&=&\frac{1}{\sqrt{2}}\left(W^6_\mu+iW^7_\mu\right), \\
Y^{--}_\mu&=&\frac{1}{\sqrt{2}}\left(W^4_\mu+iW^5_\mu\right).
\end{eqnarray}
 There are also a massive extra neutral gauge boson, $Z^\prime_\mu$, and a massless gauge boson,
$B_\mu$, which are given in terms of $W^8_\mu$ and $V_\mu$. While $B_\mu$ corresponds to the
$U_Y(1)$ gauge field, the massless fields associated with the
unbroken generators of ${SU_L(3)}$, $W^i_\mu$ ($i=1,2,3$),  turn out to be the gauge  fields of the
$SU_L(2)$ group.

At the Fermi
scale, the SM gauge group is
spontaneously broken down into the electromagnetic group via the VEVs  of the $SU_L(2) $ doublets,
$<\Phi^0_i>_0=(0,v_i/\sqrt{2})^T$ ($i=1,2$). The   SM charged gauge bosons, $W^\pm_\mu=(W^1_\mu\mp i
W^2_\mu)/\sqrt{2}$, get their masses and the bileptons receive additional mass
contributions. Finally, the $W^3$, $W^8$ and $V$ gauge fields define three neutral fields  as
follows

\begin{equation}
\left(
\begin{array}{ccc} W^3_\mu \\ W^8_\mu \\ V_\mu \end{array}
\right)=\left(
\begin{array}{ccc}
 s_W & c_W & 0 \\
 \sqrt{3} s_W & -\sqrt{3} s_W t_W &
   -\frac{1}{c_W}\sqrt{1-4 s_W^2} \\
 \sqrt{1-4 s_W^2} & -\sqrt{1-4 s_W^2} t_W &
   \sqrt{3} t_W
\end{array}
\right)\left( \begin{array}{ccc} A_\mu \\ Z_\mu \\ Z'_\mu \end{array}
\right),
\end{equation}
where $A_\mu$ corresponds to the photon, but $Z_\mu$ and $Z'_\mu$ need to be rotated to obtain the
mass
eigenstates: the SM neutral weak gauge boson $Z_1= Z \cos\theta -Z' \sin\theta $ and the extra
neutral gauge boson $Z_2=Z \sin\theta +Z' \cos\theta $, with the mixing angle $\theta$ defined by
$\sin^2\theta=\left(m^2_Z-m^2_{Z_1}\right)/\left(m^2_{Z_2}-m^2_{Z_1}\right)$. Since $\theta$ is
strongly constrained by experimental data,
we will assume that $\theta\simeq 0$ and thus the $Z$ and $Z'$ gauge bosons will be taken as the
mass eigenstates.
The masses of the heavy physical states are thus
\begin{eqnarray}
\label{mY}
m^2_{Y^{--}}&=&\frac{g^2}{4}(u^2+v^2_2+4v_3^2),\\
\label{mX}
m^2_{Y^{-}}&=&\frac{g^2}{4}(u^2+v^2_1+v_3^2),\\
\label{mZ'}
m_{Z'}^2&=&\frac{g^2}{3(1-4s_W^2)}\left(c_W^2
u^2+\frac{(1-4s_W^2)^2}{4c_W^2}\left(v_1^2+v_2^2+v_3^2\right)+3s_W^2 v_2^2\right),\\
\end{eqnarray}
where $v_3$ is the VEV of the $\Phi_3$ doublet, which is required to endow the leptons with masses.
From the symmetry breaking  hierarchy, $u>v_{1,2}>v_3$, it turns out that  $m_{Z'}> m_{Y^-,Y^{--}}>m_{W,Z}$.
In fact, neglecting the splitting between the bilepton masses,
$m_{Y^{--}}\simeq m_{Y^-}\equiv m_Y$, we obtain the following approximate relation \cite{Ng:1992st}:

\begin{equation}
m_{Y}\simeq \sqrt{\frac{3}{4}}\frac{\sqrt{1-4s_W^2}}{c_W} m_{Z'}\simeq \frac{1}{3}m_{Z'}.
\label{mZpmYrel}
\end{equation}


After SSB and once the gauge eigenstates are rotated to mass eigenstates,
the  Yang-Mills Lagrangian for the fields $W^\mu$ and $V^\mu$ can be decomposed into the SM
Yang-Mills Lagrangian plus a term that contains the interactions between the SM gauge bosons and
the heavy charged gauge bosons, together with a term that only contains interactions between the
$Z^\prime$
gauge
boson and the bileptons.  The term necessary for our calculation  can be written as

\begin{align}
\label{LSM-331}
 {\cal L}^{\rm SM-331}=&-\frac{1}{2}\left(D_\mu
Y_\nu-D_\nu Y_\mu\right)^\dag \left(D^\mu Y^\nu-D^\nu Y^\mu\right) -Y^{\dag
\mu}\left(i\,g\,{\bf W}_{\mu \nu}+i\,g'\,{\bf B}_{\mu \nu}\right)Y^\nu,
\end{align}

\noindent where  ${\bf W}_{\mu
\nu}=\tau^i \, W^i_{\mu \nu}/2$, ${\bf B}_{\mu \nu}=Y\,B_{\mu
\nu}$/2 and  $Y_{\mu}=(Y^{--}_\mu,Y^-_\mu)^T$; also, $D_\mu=\partial_\mu -ig{\bf W}_\mu -ig'{\bf
B}_\mu$ is the covariant derivative associated with the
electroweak group. From here we can get the interactions of
the bilepton gauge bosons with the photon.

As to the neutral and charged currents mediated by the heavy gauge bosons, they arise from the
fermion kinetic terms and can be written as:

\begin{equation}
 {\cal L}^{\bar{q}'q'V}=-
\frac{g}{2c_W}\left(\sum_{i=1}^3\bar{Q'}_{Li}\gamma^\mu {\bf H}_\mu Q'_{Li}+\sum_{i=1}^9 6
s_W^2 Z'_\mu \bar{q}^{'}
_{Ri}\gamma^\mu X q_{Ri}^{'}\right),
\label{lagqqV}
\end{equation}
with
\begin{eqnarray}
{\bf H}_\mu=\left(
\begin{array}{ccc}
\left(\frac{2(3X+2)s_W^2-1}{\sqrt{3}\sqrt{1-4 s_W^2}}\right)Z'_\mu & 0 &
\sqrt{2}c_W Y^{--}_\mu \\
0&  \left(\frac{2(3X+2)s_W^2-1}{\sqrt{3}\sqrt{1-4 s_W^2}}\right)Z'_\mu & \sqrt{2}c_W Y^{-}_\mu \\
\sqrt{2}c_W Y^{++}_\mu & \sqrt{2}c_W Y^{+}_\mu &  2\left(\frac{(3X-4)s_W^2+1}{\sqrt{3}\sqrt{1-4
s_W^2}}\right)Z'_\mu
\end{array}
\right),
\end{eqnarray}
where $Q'_i$ ($i=1,2,3$) is a quark triplet and
$q'_i$  is a  quark singlet, both in the flavor basis. Since the third family
has a different representation under $SU_L(3)$, after the flavor eigenstates are rotated to the mass
eigenstates there emerge FCNC couplings mediated by the $Z'$ gauge boson.
The flavor conserving $Z'$ couplings to a quark pair have the form

\begin{equation}
{\cal L}^{\bar{q} q Z'}=-\frac{g}{c_W}Z'_\mu \bar{q} \gamma^\mu\left(g_L^{'q}
P_L+g_R^{'q}P_R\right) q,
\label{lagqqZprime}
\end{equation}
where $P_{L,R}$ are the chiral projection operators and the $g^{'q}_{L,R}$ constants are presented
in Appendix \ref{FeynmanRules331}. On the other hand, the flavor-changing neutral and
charged currents  required by our
calculation can be arranged as \cite{Liu:1994rx}:

\begin{eqnarray}
{\cal L}^{NC}=-\frac{gc_W }{\sqrt{3}\sqrt{1-4s_W^2}}Z'_{\mu} U^*_{L3i}U_{L3j}
\bar{u}_i\gamma^\mu P_L  u_j ,
\label{lagYNC}
\end{eqnarray}

\begin{eqnarray}
{\cal L}^{CC}=-\frac{g}{\sqrt{2}}\left( Y^-_{\mu}U^*_{Li3}\bar{u_i}\gamma^\mu P_L
T+Y^{--}_\mu\left({U}_{Li1} \bar{D}\gamma^\mu P_L  u_i+U_{Li2} \bar{S}\gamma^\mu P_L
u_i\right)\right)+{\rm H.c.},
\label{lagYCC}
\end{eqnarray}
where $U_L$, which is the $3\times 3$ matrix that
diagonalizes the SM up quarks
from flavor eigenstates to mass eigenstates, is related to the CKM matrix by
$U_{CKM}=U_L^\dagger V_L$, with $V_L$ the mass matrix that diagonalizes the SM down quarks. Notice
that the $D_{1,2}$ flavor eigenstates can be chosen as the
mass eigenstates since the two first fermion families transform symmetrically \cite{Liu:1994rx}.

Finally, we will discuss briefly about the Yukawa couplings associated with the quark sector, which
can be written in terms of the SM quark doublets $q'_i=(u'_i,d'_i)^T$ as \cite{Liu:1994rx}:
\begin{eqnarray}
\label{LYukawadoub}
 -{\cal L}&=&\sum_{k=1}^3\sum_{i=1}^2\left(\bar{q}'_{L\,i} h^{ik}_d d'_{R\,k}\Phi_1
+\bar{q}'_{L\,i} h^{ik}_u u'_{R\,k}\tilde\Phi_2\right)+\sum_{k=1}^3\left(\bar{q}'_{L\,3} h^{3k}_d
d'_{R\,k}\Phi_2+\bar{q}'_{L\,3} h^{3k}_u u'_{R\,k}\tilde\Phi_1\right)
\nonumber\\&+&
\sum_{k=1}^3\sum_{i=1}^2\left(\bar{D}'_{L\,i} h^{ik}_d d'_{R\,k}\Delta^-
+\bar{D}'_{L\,i} h^{ik}_u u'_{R\,k}\rho^{--}\right)+\sum_{k=1}^3\left(\bar{T}_{L} h^{3k}_d
d'_{R\,k}\rho^{++}-\bar{T}_{L} h^{3k}_u u'_{R\,k}\Delta^+\right)
\nonumber\\&+&\sum_{i=1}^2\sum_{j=1}^2\left(\bar{D}'_{L\,i} h^{ij}_D D'_{R\,j}\phi^0+\bar{q}'_{L\,i}
h^{ij}_D D'_{R\,j}\phi_Y\right)+\bar{T}_{L}
h_T T_{R}\phi^{0*}-\bar{q'}_{L\,3}
h_T T_{R}\tilde\phi_Y
+{\rm H.c.},
\end{eqnarray}
where $h^{ij}$ are symmetric matrices in flavor space. After the first stage of SSB, there are
two-Higgs
doublets plus one neutral, one singly charged, and one doubly charged scalar bosons. There
will be additional scalar multiplets, which arise from the scalar sextet, that  do not couple to
the quarks. We can observe from the last line of Eq. (\ref{LYukawadoub}) that after the $\phi_Y$
doublet develops a VEV, the exotic quarks get their masses, which are thus of the order of $u$.
Furthermore, after SSB and once the mass eigenstates are obtained, there is a plethora of physical
Higgs bosons (5 neutral CP-even, 3 neutral CP-odd, 4 singly charged, and 3 doubly
charged scalar bosons), which are a mix of the Higgs eigenstates
\cite{Tully:1998wa,Tonasse:1996cx,Nguyen:1998ui}. Since these physical Higgs bosons can induce
flavor change, they will contribute to the decay $t\to c\gamma$. In particular there could be flavor
change
mediated by neutral scalar Higgs bosons due to the asymmetry in the $SU(3)$ representation of the
quark families. However, the introduction of an ad hoc discrete symmetry can eliminate any dangerous
FCNC. Since a complete treatment of the scalar sector is rather
complicated and requires to consider several parameters, we will take instead a more practical
approach: we will consider a dimension-four effective Lagrangian for typical neutral and charged
scalars that can induce the $t\to c\gamma$ decay. For the effective couplings we can consider  the
so-called Cheng-Sher ansatz \cite{Cheng:1987rs},
which is suited for models with multiple Higgs doublets. This will be useful to estimate the size of
the potential contributions to the $t\to c\gamma$ branching ratio in 331 models.

\section{Decay $t\to c\gamma$ in 331 models}
\label{calculation}

We find it useful to present our results  in a model-independent fashion. We thus consider the
following renormalizable interactions that can induce the $t\to c\gamma$ decay.  We start with the
interactions between a
neutral Higgs boson $\phi$ and a quark
pair:

\begin{eqnarray}
{\cal L}^{\phi}=-\frac{g}{c_W}\bar{q}_i\left(L^{ij}_{\phi} P_L+R^{ij}_{\phi} P_R \right)q_j
{\phi},
\label{neuscal}
\end{eqnarray}
where  $i,j=1,2,3$ stand for the quark flavors, while   $L^{ij}_{\phi}$ and $R^{ij}_{\phi}$ are  coupling
constants. If CP is conserved then $L^{ij}_{\phi}=R^{*ji}_{\phi}$. From now on, unless stated otherwise, $\phi$ will denote a neutral scalar boson. As for the singly and doubly
charged scalars, $\phi^{-}$ and $\phi^{--}$, their
interactions with SM up quarks and the exotic quarks, $D_i$ and $T$, can be
expressed in the form
\begin{eqnarray}
{\cal L}^{SCC}&=&-\frac{g}{c_W}\bar{u}_i\left(L^{iT}_{\phi^{-}} P_L + R^{iT}_{\phi^{-}} P_R
\right)T \phi^{-}-\frac{g}{c_W}\sum_{i=1,2}\bar{D}_i\left(L^{ij}_{\phi^{--}} P_L + R^{ij}_{\phi^{--}} P_R
\right) u_j \phi^{--}  +{\rm
H.c.}
\label{scalcurr}
\end{eqnarray}
As far as the gauge sector is concerned,  the most general renormalizable interactions of a neutral
gauge boson $Z'$ with a quark pair can be written as:

\begin{eqnarray}
{\cal L}^{Z'}=-\frac{g}{c_W}\bar{q}_i\gamma^\mu\left(L^{ij}_{Z'} P_L+R^{ij}_{Z'} P_R \right) q_j
Z'_\mu.
\label{neucurr}
\end{eqnarray}
Finally, the interactions of the singly and doubly charged gauge bosons to SM
and exotic quark are:

\begin{eqnarray}
{\cal L}^{GCC}=-\frac{g}{c_W}\bar{u}_i\gamma^\mu\left(L^{iT}_{Y^-} P_L + R^{iT}_{Y^-} P_R
\right)T
Y^{-}_\mu -\frac{g}{c_W}\sum_{i=1,2}\bar{D}_i\gamma^\mu\left(L^{ij}_{Y^{--}} P_L + R^{ij}_{Y^{--}} P_R \right)
u_j
Y^{--}_\mu +{\rm
H.c.}
\label{charcurr}
\end{eqnarray}
We also need  the interactions with the photon, which are dictated by electrodynamics and follow
from Eq. (\ref{LSM-331}) and the kinetic term of the scalar mutiplets. These interactions and all
the Feynman rules
necessary for our calculation are presented in Appendix \ref{FeynmanRules331}.

Due to electromagnetic gauge invariance, the $t(p_1)\to c(p_2)\gamma(q)$ decay
amplitude can be cast in the form:

\begin{equation}
{\cal M}(t\to c\gamma)
=\frac{i}{m_t}\bar{c}(p_2)\sigma_{\mu\nu} \left(
C_L P_L   +C_R P_R \right) t(p_1)q^\nu\epsilon^\mu(q).
\label{amplitude}
\end{equation}

\begin{figure}[!ht]
\begin{center}
\includegraphics[width=4in]{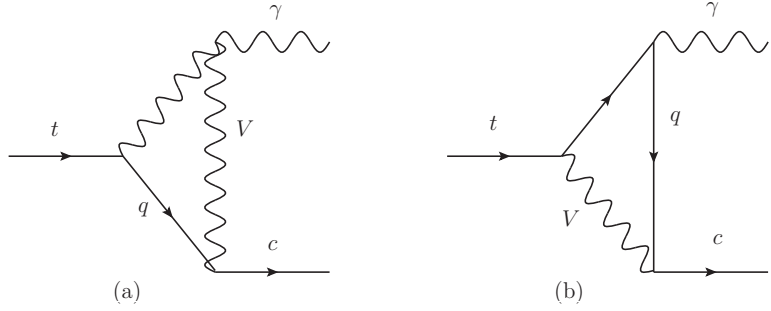}
\end{center}
\caption{Feynman diagrams, in the unitary gauge, for the decay $t\to c\gamma$ in 331 models. We show
the
contribution of an arbitrarily charged gauge boson $V$ along with an exotic or SM-like quark. The
diagram (a) does not contribute in the case of the neutral gauge boson. The contribution of charged
scalar particles are similar but with the gauge boson replaced by the scalar boson. There are also bubble diagrams that do not contribute to the $t\to c\gamma$  amplitude but are necessary to render electromagnetic gauge invariance and cancel out ultraviolet divergences.
\label{Triangles}}
\end{figure}

We show in Fig. \ref{Triangles} the one-loop contributions to the $t\to c\gamma$ decay from an
arbitrarily
charged gauge boson $V$ and a SM or exotic quark. We are using the unitary gauge, so there are no
contributions from nonphysical particles.  There are also bubble diagrams that can contribute to the
on-shell
$\bar{c}t\gamma$ vertex. Although such diagrams do not contribute to
the dipole coefficients $C_{L,R}$, they give ultraviolet divergent terms that violate
electromagnetic gauge invariance, which we have verified are canceled out by similar terms arising
from the triangle diagrams. This is similar to what happens with the $b\to s\gamma$ decay
\cite{Agrawal:1995vp,*Haisch:2008ar}. There are four possible combinations of loops carrying a quark
and a
gauge boson with the following electric charges: i) $Q_q=5/3 e$ and
$Q_V=e$, ii)$Q_q=-4/3e$ and $Q_V=-2e$, iii) $Q_q=2/3e$ and $Q_V=0$,  and iv)$Q_q=-1/3e$ and $Q_V=-e$. As
for the contribution of an arbitrarily charged scalar boson, it arises from
similar diagrams to that shown in Fig. \ref{Triangles} but with the gauge boson
replaced by a scalar boson.

 We now turn to present the results for the
contributions to the $t\to c\gamma$ decay arising from the interactions
(\ref{neuscal})-(\ref{charcurr}). In order to solve the one-loop tensor integrals, we
expressed them in terms of scalar two- and three-point scalar functions via the Passarino-Veltman
reduction scheme \cite{Passarino:1978jh}. Analytical results are given for these scalar functions in terms of dilogarithms
and  other transcendental functions.

\subsection{Gauge boson contribution}

We denote the coupling constants appearing in Eqs.
(\ref{neucurr})-(\ref{charcurr}) by  $L_V^{ij}$ and $R_V^{ij}$. After a lengthy algebra, we obtain
the contribution  from the loops carrying the arbitrarily charged gauge boson $V$ and the quark $q$:

\begin{eqnarray}
C_R^{V}&=&\frac{3eg^2}{16\pi^2c_W^2}\frac{1}{2 \delta _{{tc}}}\Bigg(
\tilde\Delta _{{qt}}\sqrt{x_t}
\left( \left(\sqrt{x_c} \left(\delta_{{qt}}+2\right) R_V^{{tq}}+ \sqrt{x_q}\delta
_{{tc}}L_V^{{tq}}\right)R_V^{{qc}}-\sqrt{x_t}\left(\delta _{{qc}}+2\right) L_V^{{qc}}
L_V^{{tq}}\right)
\nonumber\\&-&\frac{Q_t\delta_q}{ \sqrt{x_c} } \left(
   \sqrt{x_c} \left(\delta
   _{{qt}}+2\right) L_V^{{qc}}L_V^{{tq}}+\sqrt{x_q}  \delta
   _{{tc}}L_V^{{qc}}R_V^{{tq}}- \sqrt{x_t}\left(\delta
   _{{qc}}+2\right) R_V^{{tq}}R_V^{{qc}} \right)G_{(c,q,V)}
\nonumber\\&+&\frac{2Q_q}{\sqrt{x_q}}\Big(\sqrt{x_c x_t  x_q}  \left(\delta
   _{{qc}}+2\right) R_V^{{tq}}R_V^{{qc}} -
 x_t \left(\sqrt{x_c}  \delta
   _{{tc}}R_V^{{tq}}+ \sqrt{x_q} \left(\delta
   _{{qt}}+2\right)L_V^{{tq}}\right)L_V^{{qc}}\Big)H_{(c,t,q,V)}
\nonumber\\&+&2\Delta _{{qt}}\Big(\sqrt{x_t} \left(\sqrt{x_c}
    \left(\delta_{qc}+\delta_{tc}+2\right)R_V^{{tq}}-3
   \sqrt{x_q} \delta _{{tc}}L_V^{{tq}}\right)R_V^{{qc}}- x_t \left(\delta_{qt}-
   \delta_{tc}+2\right)L_V^{{qc}}
   L_V^{{tq}}\Big)H_{(c,t,V,q)}\nonumber\\&+&\frac{1}{ \delta _{{tc}}}\Big[
 x_t \sqrt{x_c}
   \left(Q_t \left(x_c
   \left(x_q-3\right)+\delta_q
   \left(x_q+2\right)\right)-x_t \left(\delta _{{qc}} \tilde\Delta
   _{{qt}}+4 \Delta _{{qt}}\right)\right)R_V^{{qc}}R_V^{{tq}}\nonumber\\&+&
   \Big(Q_t\left(2 x_t \left(x_q^2-\left(1-\delta_q\right) \delta_t\right)-x_c
   \left(x_q
\left(\delta_{qt}-\delta_t\right)-\delta_t-1\right)\right)+2\Delta_{qt}x_c
\left(x_q+2\right) x_t-\tilde\Delta_{qc}x_c x_t^2\Big)L_V^{{qc}}L_V^{{tq}}\nonumber\\&-& \sqrt{x_c
x_q}
    \delta _{{tc}} \left(x_t \left(\Delta
   _{{qt}}-3 Q_q\right)-Q_t\delta_q
   \right)L_V^{{qc}}R_V^{{tq}} -6\sqrt{x_tx_q}
    Q_t \delta _{{tc}}R_V^{{qc}}L_V^{{tq}}\Big]F_{(c,t,q,V)}
  \Bigg),
\label{CRV}
\end{eqnarray}
and
\begin{eqnarray}
C_L^{V}&=&C_R^{V}\left(t\leftrightarrow c\right),
\label{CLV}
\end{eqnarray}
where $\Delta_{ij}=Q_i-Q_j$ and $\tilde\Delta_{ij}=2Q_i-Q_j$, with $Q_i$ the electric charge of
particle $i$ in units of $e$; we also introduced the definitions $x_i=m_i^2/m_V^2$, $\delta_{ij}=x_j-x_i$ and
$\delta_{i}=1-x_i$. The $F_{(i,j,k,l)}$, $G_{(i,j,k)}$ and
$H_{(i,j,k,l)}$ functions are given in terms of Passarino-Veltman  scalar functions:

\begin{eqnarray}
\label{PassVelFun1}
F_{(i,j,k,l)}&=&B_0(m_i^2,m_k^2,m_l^2)-B_0(m_j^2,m_k^2,m_l^2),\\
\label{PassVelFu2}
G_{(i,j,k)}&=&B_0(0,m_j^2,m_k^2)-B_0(m_i^2,m_j^2,m_k^2),\\
\label{PassVelFun3}
H_{(i,j,k,l)}&=&m_k^2 \,C_0(m_i^2,m_j^2,0,m_k^2,m_l^2,m_k^2),
\end{eqnarray}
where $B_0$  and $C_0$ are two-  and three-point scalar functions written in the notation of Ref. \cite{Mertig:1990an}. From here it is clear that
ultraviolet divergences cancel out.
Explicit integration of the above functions yields

\begin{eqnarray}
\label{Fct}
F_{(i,j,k,l)}&=&f(m_i^2,m_j^2,m_k^2,m_l^2)+\frac{1}{m_i^2}\left(f_{-}
-f_{+}\right)\left(m_i^2,m_k^2,
m_l^2\right)-\frac{1}{m_j^2}\left(f_{-} -f_{+}\right)\left(m_j^2,m_k^2,m_l^2\right),\\
\label{Gc}
G_{(i,j,k)}&=&g(m_i^2,m_j^2,m_k^2)+\frac{1}{m_i^2}\left(f_{+}
-f_{-}\right)\left(m_i^2,m_j^2,mk^2\right),\\
\label{HctqV}
H_{(i,j,k,l)}&=&\frac{m_k^2}{m_j^2-mi^2}\left(\left(h_{+} +h_{-}\right)\left(m_i^2,m_k^2,
m_l^2\right)-\left(h_{+} +h_{-}\right)\left(m_j^2,m_k^2,m_l^2\right)\right),
\end{eqnarray}
with

\begin{eqnarray}
f(x,y,w,z)&=&\frac{(x-y)(w-z)}{2 x y}
   \log \left(\frac{w}{z}\right),\\
   f_\pm(x,y,z)&=&\sqrt{\lambda(x,y,z)}\,{\rm arctanh}\left(\frac{y-z\pm
x}{\sqrt{\lambda(x,y,z)}}\right),\\
g(x,y,z)&=&\frac{(y-z)^2-x(y+z)}{2x(y-z)}\log\left(\frac{y}{z}\right)-1,\\
h_{\pm}(x,y,z)&=&{\rm Li}_2\left(\frac{2x}{x+y-z\pm \sqrt{\lambda(x,y,z)}}
\right),
\end{eqnarray}
and $\lambda(x,y,z)=\left(x-y-z\right)^2-4 y z$.

A special case arises when only the left-handed quarks interact with the exchanged gauge boson, as
occurs with the bilepton contributions in the minimal 331 model. In this scenario we have

\begin{eqnarray}
C_R^{Y}&=&\frac{3eg^2}{16\pi^2c_W^2}\frac{L_Y^{{tq}}L_Y^{{qc}} }{2 \delta _{{tc}}}\Bigg(
 \tilde\Delta _{{qt}}x_t\left(\delta _{{qc}}+2\right) +\frac{1}{ \sqrt{x_c} }Q_t\delta_q \left(
   \sqrt{x_c} \delta
   _{{qt}}+2\right)  G_{(c,q,Y)}+2Q_q
 x_t  \left(\delta
   _{{qt}}+2\right)H_{(c,t,q,Y)}\nonumber\\&-&2\Delta _{{qt}}x_t \left(\delta_{qt}-
   \delta_{tc}+2\right)H_{(c,t,Y,q)}-\frac{1}{ \delta _{{tc}}}
    \Big(Q_t\left(2 x_t \left(x_q^2-\left(1-\delta_q\right) \delta_t\right)-x_c
   \left(x_q
\left(\delta_{qt}-\delta_t\right)-\delta_t-1\right)\right)\nonumber\\&+&2\Delta_{qt}x_c
\left(x_q+2\right) x_t-\tilde\Delta_{qc}x_c x_t^2\Big)F_{(c,t,q,Y)}
  \Bigg).
\label{CRY331}
\end{eqnarray}

\subsection{Scalar contribution}
We now consider the contribution from the loops carrying an arbitrarily charged scalar boson $\phi$
and a quark $q$, which arise from Feynman diagrams similar to those of Fig. \ref{Triangles} but with
the gauge boson replace by a scalar boson.  The Passarino-Veltman
reduction scheme yields the following
$C_{L,R}^\phi$ coefficients:

\begin{eqnarray}
C_R^{\phi}&=&\frac{3eg^2}{16\pi^2c_W^2} \frac{1}{2\eta_{tc}}\Bigg(\frac{\sqrt{y_t}}{\sqrt{y_q}}
Q_q
   \left(\sqrt{y_q}\left(\sqrt{y_c} L_{\phi}^{{qc}} R_{\phi}^{{tq}}-\sqrt{y_t}
L_{\phi}^{{tq}}R_{\phi}^{{qc}}\right)-\eta_{{tc}}R_{\phi}^{{tq}}R_{\phi}^{{qc}}
\right)H_{(c,t,q,\phi)}
\nonumber\\&+&2\sqrt{y_t}
   \left(\sqrt{y_c}
   L_{\phi}^{{qc}} R_{\phi}^{{tq}} -\sqrt{y_t}L_{\phi}^{{tq}}
R_{\phi}^{{qc}}\right)\left(\tilde\Delta_{{qt}}+\Delta_{{qt}}H_{(c,t,\phi,q)}\right)+\frac{1}{\sqrt{
y_c }}Q_t
\eta_q \left(\sqrt{y_c} L_{\phi}^{{tq}}
   R_{\phi}^{{qc}}-\sqrt{y_t}L_{\phi}^{{qc}} R_{\phi}^{{tq}}
\right)G_{(c,q,\phi)}\nonumber\\&+&\frac{1}{\eta
_{{tc}}}\Big(\left( y_c y_t \tilde\Delta_{{qt}}+{Q_t}\eta_q (y_c-2y_t )\right)L_{\phi}^{{tq}}
R_{\phi}^{{qc}}+
\sqrt{y_t}
   \left(\sqrt{y_c} \left({Q_t}\eta_q-y_t \tilde\Delta
   _{{qt}}\right) L_{\phi}^{{qc}}+2 {Q_t} \sqrt{y_q}
    \eta _{{tc}}R_{\phi}^{{qc}}\right)R_{\phi}^{{tq}}\Big)F_{(c,t,q,\phi)}
  \Bigg),\nonumber\\
\label{CRS}
\end{eqnarray}
where $y_i=m_i^2/m_\phi^2$, $\eta_{ij}=y_i-y_j$, and $\eta_i=1-y_i$. $C_L^\phi$ can be obtained from
$C_R^\phi$ after the replacements $t\leftrightarrow c$,
$L^{tq}_\phi \leftrightarrow R^{qc}_\phi$, and $R^{tq}_\phi \leftrightarrow L^{qc}_\phi$  are done:

\begin{eqnarray}
C_L^{\phi}&=&C_R^{\phi}\left(\begin{array}{c}t\leftrightarrow c\\ L \leftrightarrow
R\end{array}\right).
\label{CLS}
\end{eqnarray}
For a neutral CP-even neutral scalar boson, denoted by $\phi$ rather than $\phi^0$ to avoid to be plagued by indices, $R^{ij}_\phi=L^{ij}_\phi\equiv \lambda^{ij}_\phi$,
the above expression reduces to

\begin{eqnarray}
C_R^{\phi}&=&\frac{eg^2}{16\pi^2c_W^2}  \frac{\sqrt{y_t}\lambda^{{qc}}_\phi
\lambda^{{tq}}_\phi}{\left(\sqrt{y_c}+\sqrt{y_t}\right)}
\Bigg(1+\frac{1}{\sqrt{y_c}}  \eta_q G_{(c,q,\phi)}+\frac{2}{\sqrt{y_q}}
   \left(\sqrt{y_q}+\sqrt{y_c}+\sqrt{y_t} \right)H_{(c,t,q,\phi)}
\nonumber\\&+&\frac{1}{\eta _{{tc}}}\left( \sqrt{y_c y_t}  -2
\sqrt{y_q}(\sqrt{y_c}+\sqrt{y_t})+\eta_q\left(2+\sqrt\frac{{y_c}}{{y_t}}\right)
    \right)F_{(c,t,q,\phi)}
  \Bigg).\nonumber\\
\label{CRSneut}
\end{eqnarray}

\section{Numerical analysis and discussion}
\label{analysis}

In addition to the SM contribution to the $t\to c\gamma$ decay, the new contribution
from the gauge sector of the minimal 331 model can be
written as

\begin{equation}
C_{L,R}^G=\sum_{q=u,c,t}C_{L,R}^{Z'}+\sum_{q=T}C_{L,R}^{Y^-}+\sum_{q=D,S}C_{L,R}^{Y^{--}},
\label{CLRG}
\end{equation}
while the contribution from the Higgs sector is as follows

\begin{equation}
C_{L,R}^H=\sum_{q=u,c,t}C_{L,R}^{\phi}+\sum_{q=T}C_{L,R}^{\phi^-}+\sum_{q=D,S}C_{L,R}^{\phi^{--}
}, \label{CLRS}
\end{equation}
where it is assumed that we must consider all the physical neutral and charged Higgs bosons.  In the
case of 331 models without exotic quarks, the generic contribution arises only from the $Z'$ gauge
boson and the neutral scalar bosons.

From Eq. (\ref{amplitude}), the corresponding $t\to c\gamma$ decay width follows easily:

\begin{equation}
\label{decaywidth}
\Gamma(t\to c \gamma)=\frac{m_t}{16\pi}\left(1-\frac{m_c^2}{m_t^2}\right)^3
\left(\left|{C}_L^G+C_L^H\right|^2+\left|{C}_R^G+C_R^H\right|^2\right).
\end{equation}
In order to get a realistic  estimate for $\Gamma(t\to c\gamma)$, we will consider the current
constraints on the masses of the heavy gauge boson, the exotic quarks, and the scalar bosons.

\subsection{Constraints on the model parameters}

Considerable work has gone into studying constraints on the masses of the extra gauge bosons  of 331
models. The most stringent bound on the mass of a doubly charged bilepton was obtained from
muonium-antimuonium conversion \cite{Willmann:1998gd}. This bound, $m_{Y^{--}}>800$ GeV,  is based
on the assumptions that the bilepton-lepton couplings are flavor diagonal and the scalar sector of
the model does not contribute significantly to muonium-antimuonium conversion. Another stringent
bound, $m_{Y^{--}}>750$ GeV, arises from fermion pair production and lepton-flavor violating
processes \cite{Tully:1999yg}. It has been argued \cite{Pleitez:1999ix}, however, that these bounds
can be evaded if one makes less restrictive assumptions than  the aforementioned
analyses. As for the $Z'$ gauge boson mass, it is related to the  bilepton masses by
Eq. (\ref{mZpmYrel}): $m_{Z'}\simeq 3 m_Y$. Therefore, the most stringent bounds on
the doubly charged bilepton mass translates into a lower bound on $m_{Z^\prime}$ of about 2 TeV,
which is similar to other restrictive bounds obtained in Refs.
\cite{CarcamoHernandez:2005ka,Dias:2004dc,Ramirez-Barreto:2007mt}.

There is considerably less literature dealing with bounds on the exotic quark masses
\cite{Das:1998qh,GonzalezSprinberg:2005zd}, which in general depend on the masses of the heavy gauge
bosons. From the search for supersymmetric particles at the Tevatron, a bound on the $D$ quark mass
of about 300 GeV was obtained for $m_{Z'}$ around 1 TeV \cite{Das:1998qh}. Another constraint was
obtained in \cite{GonzalezSprinberg:2005zd} from the experimental data on the $Z\to
\bar{b}b$ decay and electroweak precision measurements at the $Z$ pole: it was found that the $T$
quark mass is bounded into the 1500--4000 GeV interval for   $m_{Y^{--}}$ around 700 GeV.

As far as the bounds on the scalar boson masses are concerned, these are more difficult to obtain as
the scalar sector  of the minimal 331 model is plagued with free parameters. There are a few recent
bounds on the charged scalar boson masses from direct searches at the LHC, but they are model
dependent.  As will be
discussed below, we will consider the scenario in which there is only a relatively light neutral
scalar boson, with a mass of a few hundreds of GeVs, whereas the remaining scalar bosons will be
assumed to be very heavy. Hence the bulk of the scalar contribution would arise from the neutral
scalar boson.

In conclusion, in our analysis below we will consider degenerate bileptons with a mass above 600
GeV, whereas for the extra neutral gauge boson mass we will assume the relation $m_{Z'}\simeq 3 m_Y$. For the exotic quarks we will assume the hierarchy $m_T\sim m_S>m_D$, with $m_D$,
and $m_S$ around 500 GeV and 1000 GeV, respectively. Furthermore, the existence of a
relatively light neutral Higgs boson with FCNC couplings and a mass of a
few hundreds of GeVs, will be assumed.

\subsection{Gauge boson contribution}

\begin{figure}[!ht]
\begin{center}
\includegraphics[width=4in]{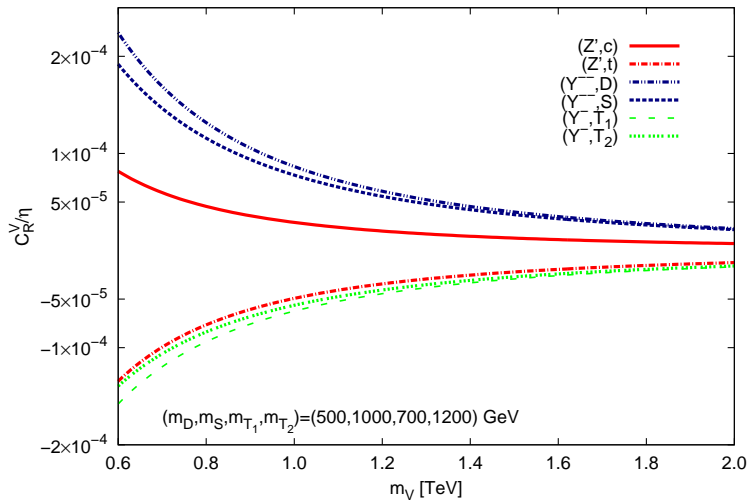}
\end{center}
\caption{Partial contribution from the heavy gauge boson  and the internal  quark pair ($V,q$) to
$C_R^V$  as a function of $m_V$ and for fixed values of the exotic quark masses in the minimal 331
model. We considered two values for the mass
of the $T$ quark ($m_{T_1}$ and $m_{T_2}$). Also
 $\eta=U_{L21}U^*_{L31}$ for $(Y^{--},D)$, $\eta=U_{L22}U^*_{L32}$ for $(Y^{--},S)$,
and $\eta=U_{L23}U^*_{L33}$ for $(Z',c)$, $(Z',t)$ and $(Y^-,T)$.\label{CRVmV}}
\end{figure}

In order to estimate the size of  the $t\to c\gamma$ decay width, we show in Fig. \ref{CRVmV}  the
behavior of
the partial contributions from the heavy gauge bosons to the $C_R^V$ coefficients as  a function of
the gauge boson and the exotic quark masses.  We do not show the $C_L^V$ coefficient as its size is more than two orders of magnitude below than that of $C_R^V$ due to the small value of the $c$ mass.
Each contribution
was divided by the associated products of $U_L$ matrix elements, which are encapsulated in the
$\eta$ coefficient.  A word of caution is in order here: the values shown in Fig. \ref{CRVmV} can be
dramatically reduced if $\eta$ is much smaller than unity. In the case of 
the extra neutral gauge boson $Z'$, we only show the contribution from the loops carrying the $c$
and $t$ quarks since the amplitude corresponding to
the  $u$ quark involves two flavor-changing vertices and it is expected to be more suppressed. From
Fig. \ref{CRVmV}, we can conclude that the largest contribution to $C_R^V$ could arise from the
charged bileptons, while the smallest contribution could be due to the extra neutral gauge boson.
Another point worth mentioning is that, while the $C_R^V$ coefficient is strongly dependent on the
value of the gauge boson mass and can decrease up to one order of magnitude in the
interval from 600 GeV to 2000 GeV, its change is almost imperceptible when the
exotic quark mass is varied in a similar interval, which is evident in Fig. \ref{CRVmV} as the
curves corresponding to  exotic quarks with same electric charges but distinct masses almost
overlap. Although there could  be large cancellations when summing over all the contributions to
$C_R^V$, it is interesting to point out  that a mechanism such as the one that  suppresses
the contribution from the $W$ gauge boson does not  operate in the case of the charged bileptons,
even if we assume that they are mass degenerate. In such a case, from Eq. (\ref{CRY331})
we can see that $C_R^Y$
would adopt the form

\begin{equation}
C_R^Y=\sum_{i=1}^3 U_{L2i}U^*_{L3i}f(m_{Q_i},m_Y,Q_Y),
\end{equation}
where the sum runs over the exotic quarks. Here $m_Y$  stands for the bilepton mass, while the
bilepton electric charge is  $Q_Y=Q_i-Q_t$. Since $U_L$ is unitary, $C_R^Y$ would vanish if the $f$
function  was independent of  $m_{Q_i}$ and $Q_i$. However, even if the exotic quarks  were mass
degenerate, $C_R^V$ would not vanish as they do not have the same electric charge.
Therefore, we do not expect a strong suppression of  $C_R^Y$. In the case of the $Z'$ contribution,
we also do not expect large cancellations between the $c$ and $t$ contributions
because of the disparity of the top quark mass and  the nonuniversality of the couplings of the $Z'$
to SM quarks.

\begin{figure}[!ht]
\begin{center}
\includegraphics[width=4in]{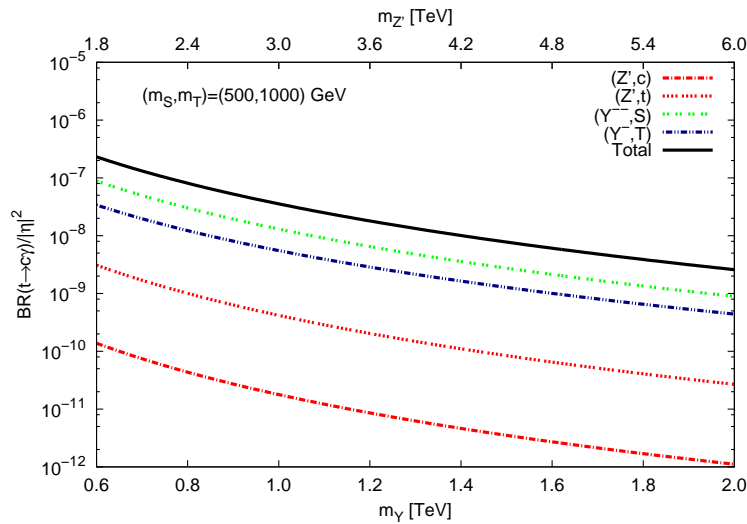}
\end{center}
\caption{Partial contributions from the heavy gauge boson  and the internal  quark  ($V,q$) to
the $t\to c\gamma$ branching ratio in the minimal 331 model as a function of the bilepton mass,
considering
degenerate bileptons, and for fixed values of the exotic quark masses. The ${Z'}$ mass is related to
$m_Y$ by Eq. (\ref{mZpmYrel}): $m_{Z'}\simeq 3 m_Y$. Also $\eta=U_{L22}U^*_{L32}$ for
$(Y^{--},S)$, and
$\eta=U_{L23}U^*_{L33}$ for $(Z',c)$, $(Z',t)$ and $(Y^-,T)$.  We also show the total contribution
to $BR(t\to c\gamma)$ considering  the scenario discussed in the text, namely, when
$U_{L21}U^*_{L31}\sim 0$ and
$U_{L22}U_{L32}^*\sim
-U_{L23}U_{L33}^*$.
\label{BVmV}}
\end{figure}

We now consider an scenario in which the heavy
gauge boson contributions to the $t\to c\gamma$ decay add up instead of canceling out. In such a case, a rough
estimate  for
the  branching ratio, $BR(t\to c\gamma)$, is that it would be of the same order of
magnitude than its partial contributions: although we would need to know the actual values of the
$U_L$ matrix elements
to obtain the total contribution,  an enhancement of several orders of magnitude with respect to
each contribution cannot be expected. We thus show the individual behavior of the partial
contributions to $BR(t\to c\gamma)$ in Fig. \ref{BVmV} as a function of the bilepton mass and
illustrative values of the exotic quark masses. Since we use the relation  $m_{Z'}\simeq 3 m_Y$, the $Z'$ contributions is considerably more suppressed as compared to the bilepton contribution. We can conclude that it would be very unlikely that
the total contribution of the heavy gauge bosons to $BR(t\to c\gamma)$ would surpass the $10^{-7}$
level even if there were no large cancellations between the
partial contributions or a further suppression coming from  the $U_L$ matrix
elements. In order to illustrate our point, we assume a simple scenario in which
$U_{L21}U^*_{L31}\sim
0$ and $U_{L22}U_{L32}^*\sim -U_{L23}U_{L33}^*$. We then calculate the total contribution
to the $t\to c\gamma$ branching ratio from the heavy gauge bosons: the result is given by the solid
line shown in the plot of Fig. \ref{BVmV}. Although the total $BR(t\to c\gamma)$  is slightly
enhanced, even
if we  consider nondegenerate bileptons, such an enhancement would hardly surpass one order of
magnitude. Finally, we also expect that the generic contribution to  $BR(t\to c\gamma)$ from the
gauge sector of  331 models  is of the order of $10^{-9}$ at most since it only arises from the
extra neutral $Z'$ gauge boson, whose mass is constrained, from experimental data, to be much larger than
1 TeV.

\subsection{Scalar boson contribution}
\begin{figure}[!ht]
\begin{center}
\includegraphics[width=4in]{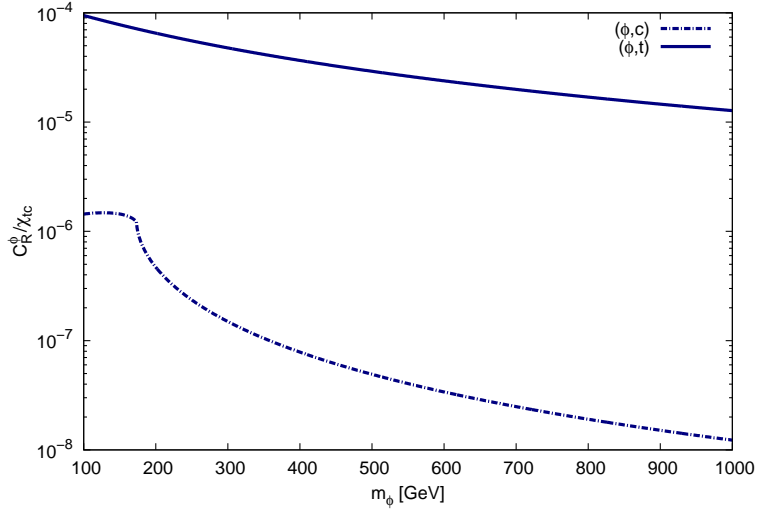}
\end{center}
\caption{Partial contribution from the loops carrying a neutral scalar boson and the $c$ or $t$ quarks
to $C_R^\phi$  as a function of the scalar boson mass. For the $\phi \bar{c}t$ coupling we used the Cheng-Sher parametrization.
\label{CRSmS}}
\end{figure}

Since there are several physical neutral, singly, and doubly charged scalars, along with several
free parameters, such as the masses of the scalar bosons, mixing angles, and Yukawa couplings, the
analysis of this contribution turns out to be very complicated. Fortunately, at low energies, as far
as the quark sector is concerned, the scalar sector of the minimal 331 model resembles that of a
two-Higgs doublet model \cite{Liu:1993gy}. Therefore, in our analysis we will assume that the largest contribution from the scalar
sector arises from the lightest neutral scalar boson. This is equivalent to assume that  the
remaining scalar bosons are very heavy or that there is a large suppression of the associated
Yukawa couplings.  Hence we will  analyze the behavior of the contribution from a typical neutral
scalar boson with a mass of a few hundreds of GeVs. For the coupling of such a scalar boson to a SM
quark pair we will consider the Cheng-Sher ansatz \cite{Cheng:1987rs}, which  is meant for mutiple-Higgs-doublets models. We will thus assume that
$\lambda_{\phi}^{ij}= \sqrt{m_im_j}\chi_{ij}/(2m_Z)$, with $\chi_{ij}$ a
number of the order of
unity at most. We are compelled to make this assumption due to our ignorance of the parameters involved in the scalar sector of the model. Although this can be a very
optimistic assumption that can led us to overestimate the scalar contribution to the $t\to c\gamma$
decay, one must have in mind that there is a suppression factor, $\chi_{ct}$,  whose value could be very suppressed.   In Fig. \ref{CRSmS} we show the
partial contribution from the neutral scalar boson
accompanied by the $c$ and $t$ quarks to the $C_R^\phi$ coefficient as a function of $m_{\phi}$.
In this case $C_L^\phi=C_R^\phi$ and we do not show the contribution of the $u$ quark   as it is
several orders of magnitude below than the $c$ quark contribution.
From this plot, we can also conclude that
the scalar contribution to the $t\to c\gamma$ decay could be of the same order of magnitude than the
gauge
boson contribution. We also note that  $C_R^\phi$ decreases rapidly as the scalar boson mass
increases, but it depends considerably on the mass of the
internal quark, which in fact is due to the use of the
Cheng-Sher parametrization. It is also worth noting that the plateau observed in the case of the $c$
quark contribution to $C_R^\phi$  is a reflect of the fact that below the mass threshold
$m_\phi=m_t-m_c$, namely $m_\phi \lesssim 174$ GeV, the $t$ quark can decay as $t\to c\phi$, and so
the Higgs-mediated $t\to c\gamma$ transition amplitude gets  enhanced. Beyond this mass threshold,
the $t\to c\phi$ decay is no longer  kinematically allowed and $C_R^\phi$ becomes more suppressed as
$m_\phi$ becomes heavier.

\begin{figure}[!ht]
\begin{center}
\includegraphics[width=4in]{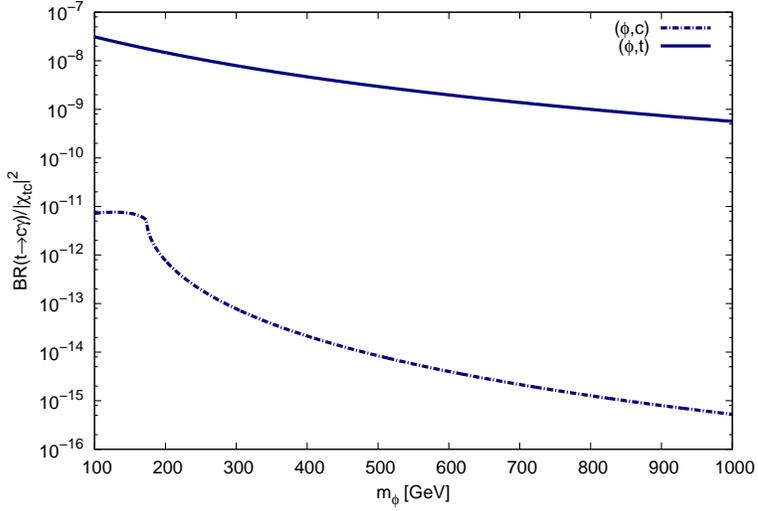}
\end{center}
\caption{Partial contributions from the loops carrying a neutral scalar boson $\phi$ and the $c$ or $t$ quarks
to
$BR(t\to c\gamma)$ as a function of the
scalar boson mass.  For the $\phi \bar{c}t$ coupling we used the Cheng-Sher parametrization.
\label{BSmS}}
\end{figure}

The individual contributions from a neutral scalar boson and the $c$ and
$t$ quarks to $BR(t\to c\gamma)$  are shown  Fig. \ref{BSmS} as functions of the scalar boson mass.
We observe that the $t$ quark contribution is much larger than that of the $c$ quark, hence  we
expect that the bulk of the scalar contribution to  $BR(t\to c\gamma)$ would arise mainly from the
loop carrying an internal top quark. Therefore the scalar contribution would be of the same order of
magnitude than the gauge boson contribution.

\section{Final remarks}
\label{conclusions}
We have analyzed the one-loop-induced decay $t\to c\gamma$ in the framework of 331 models,
with particular emphasis on the minimal version. The generic contribution of these class of models
to this decay is induced by a new neutral gauge boson and a neutral scalar boson. In the minimal
model there are also the contributions of singly and doubly charged gauge and scalar bosons,
accompanied by exotic quarks. We have found that, given the current constraints on the masses of the
new particles, the dominant contribution to the $t\to c\gamma$ branching ratio could
arise from the new charged gauge bosons and the lightest neutral scalar boson, although a
branching ratio enhancement could be expected if the remaining scalars are also relatively light and have flavor-changing couplings.
Contrary to the
case of
the contribution of the $W$ gauge boson, the bilepton gauge boson contribution is free from large
cancellations even if the bileptons  are mass degenerate: there is an imperfect
GIM-like mechanism in the minimal 331 model, which stems from the fact that the exotic quarks do not
share the same electric charge. We examined a scenario in which $BR(t\to c\gamma)$ could be of the
order of $10^{-7}$, but this value could be strongly suppressed at it has a large dependence on the
values of the mixing matrix, $U_L$, that rotates up quarks from the flavor to the mass basis. For
instance,  if the $U_L$ matrix elements are of the order of $10^{-1}$, the bilepton contribution to
the $t\to c\gamma$ branching ratio would be of the order $10^{-11}$. In order to have an estimate for the contribution of the
neutral
scalar boson, we considered the Cheng-Sher ansatz for the flavor-changing couplings of the Higgs
boson and found that the contribution to the $t\to c\gamma$ branching ratio could be of the order of
$10^{-7}$ for a Higgs boson with a mass of the order of 100--200 GeV. A point worth to mention is
that, in 331 models without exotic quarks, the main contribution could arise from the lightest
neutral scalar since the $Z'$ mass is strongly constrained and so this contribution would be of
the order of $10^{-9}$ at most.

As long as a particular 331 model was realized in nature, a more
reliable estimate of the $t\to c\gamma$ decay would be
obtained once more details of the model were known. We must conclude that any potential effects of
331 models on the $t\to c\gamma$ decay would hardly  be observed in a near future.

\acknowledgments{We would like to thank Conacyt (M\'exico) for support. Partial support from VIEP
(BUAP) is also acknowledge.}

\appendix
\section{Feynman rules for the $t\to c\gamma$ decay in the minimal 331 model}
\label{FeynmanRules331}
We first present the $L_V^{ij}$ and $R_V^{ij}$ coefficients necessary for the numerical evaluation
of the $C_{L,R}^V$ coefficients in the minimal 331 model.
The flavor conserving couplings of the $Z'$ gauge boson to SM up quarks
have the form of Eq. (\ref{lagqqZprime}). The coefficients necessary for the calculation of the
$t\to c\gamma$ amplitude can be extracted from Eq. (\ref{lagqqV}) and are given by

\begin{eqnarray}
L_{Z'}^{uu}=L_{Z'}^{cc}&=&-\frac{1-2s_W^2}{2\sqrt{3}\sqrt{1-4s_W^2}},\\
L_{Z'}^{tt}&=&\frac{1}{2\sqrt{3}\sqrt{1-4s_W^2}},\\
R_{Z'}^{uu}=R_{Z'}^{cc}=R_{Z'}^{tt}&=&\frac{2s_W^2}{\sqrt{3}\sqrt{1-4s_W^2}}.
\end{eqnarray}
On the other hand, the coupling constants for the flavor-changing  interactions
of the heavy gauge bosons  are purely left-handed, as shown in Eqs. (\ref{lagYNC}) and
(\ref{lagYCC}). The corresponding coupling constants are presented in Table
\ref{FCFeynmanRules}. Notice that the singly charged  bilepton does not couple to a pair of SM
quarks but only to a SM quark and an exotic quark.

\begingroup
\squeezetable
\begin{table}[!ht]
\begin{tabular}{ll}
\hline
Vertex ($\bar{q}_i q_j V$)  &$L_V^{ij}$\\
\hline
$\bar{u}cZ'$&$\frac{c_W^2 U^{*}_{L31}U_{L32}}{\sqrt{3}\sqrt{1-4s_W^2}}$\\
$\bar{u}tZ'$&$\frac{c_W^2 U^{*}_{L31}U_{L33}}{\sqrt{3}\sqrt{1-4s_W^2}}$\\
$\bar{c}tZ'$&$\frac{c_W^2 U^{*}_{L32}U_{L33}}{\sqrt{3}\sqrt{1-4s_W^2}}$\\
$\bar{c}TY^{-}$&$\frac{c_W U^{*}_{L23}}{\sqrt{2}}$\\
$\bar{t}TY^{-}$&$\frac{c_W U^{*}_{L33}}{\sqrt{2}}$\\
$\bar{D}cY^{--}$&$\frac{c_W U_{L21}}{\sqrt{2}}$\\
$\bar{S}cY^{--}$&$\frac{c_W U_{L22}}{\sqrt{2}}$\\
$\bar{D}tY^{--}$&$\frac{c_W U_{L31}}{\sqrt{2}}$\\
$\bar{S}tY^{--}$&$\frac{c_W U_{L32}}{\sqrt{2}}$\\
\hline
\end{tabular}
\caption{Coupling constants  for the flavor
changing vertices involving gauge
bosons in the minimal 331 model. The right-handed coupling constants
vanish and $U_L$ stands for the mixing matrix that diagonalizes the SM up quarks.
\label{FCFeynmanRules}}
\end{table}
\endgroup

For our calculation we also  need the interaction of the photon with charged particles. Apart
from the usual coupling of a photon with a fermion pair, $-ie Q_f\gamma^\mu$, the couplings of a
charged gauge boson with the photon can be extracted from Eq. (\ref{LSM-331}) and can be written as:

\begin{equation}
{\cal L}^{YYA}=i\,eQ_Y\left( A^\mu \left(Y_{\mu
\nu}Y^{\dagger \nu}-Y^\dagger_{\mu \nu}Y^{\nu}\right)-F_{\mu
\nu}Y^{\mu}Y^{\dagger\nu}\right),
\label{LYYA}
\end{equation}
with $Y_{\mu\nu}=\partial_\mu Y_\nu-\partial_\nu Y_\mu$. A similar term determines the interaction
of the $Z'$ gauge boson with the bilepton gauge bosons. As for the couplings of the photon with the
new physical charged scalar bosons, they emerge from the kinetic term of the scalar triplets and
sextet after rotating to the mass eigenstates. For a typical charged scalar boson $\phi$ we have

\begin{equation}
{\cal L}^{\phi\phi A}=\left(D^e_\mu \phi\right)^\dag
\left(D^{e\,\mu}\phi\right),
\label{LSSA}
\end{equation}
where $D^{e\,\mu}=\partial^\mu+ie Q_\phi A^\mu$ is the $U_{em}(1)$ covariant derivative.
The Feynman rules for these vertices are presented in Table \ref{QEDRules}.

\begingroup
\squeezetable
\begin{table}[!ht]
\begin{tabular}{l  l}
\hline
Vertex&Feynman rule\\
\hline
$A_\mu(k_3) Y_\alpha(k_1)Y^\dagger_\beta(k_2)$&$-ieQ_Y\Gamma_{\alpha\beta\mu}(k_1,k_2,k_3)$\\
$Z'_\mu(k_3)
Y_\alpha(k_1)Y^\dagger_\beta(k_2)$&$\frac{ig}{2c_W}\sqrt{3}\sqrt{1-4s_W^2}
\Gamma_{\alpha\beta\mu}(k_1,k_2,k_3)$\\
$A_\mu \phi\phi^\dagger$&$-iQ_\phi(k_1-k_2)_\mu$\\
\hline
\end{tabular}
\caption{Feynman rules for the electromagnetic vertices involving charged gauge and scalar bosons.
$Q_Y$ ($Q_\phi$) stands for the charge of the gauge (scalar) boson,  and
$\Gamma_{\alpha\beta\mu}(k_1,k_2,k_3)=(k_1-k_2)_\mu g_{\alpha \beta}
+(k_2-k_3)_\alpha g_{\beta\mu}+(k_3-k_1)_\beta g_{\mu\alpha}$, with all the momenta incoming.
\label{QEDRules}}
\end{table}
\endgroup

\bibliographystyle{apsrev4-1}
\bibliography{revisedarticle}{}

\end{document}